\documentclass[intlimits,twoside,a4paper]{article}

\usepackage{amsmath,amssymb}
\usepackage{graphicx}
\usepackage{epsfig}
\usepackage{epstopdf}
\usepackage[T2A]{fontenc}
\usepackage[cp1251]{inputenc}

\usepackage{cmpj2}

\issue{2015}{18}{2}{23602}
\doinumber{10.5488/CMP.18.23602}

\title[Spatial-temporal redistribution of point defects \ldots]%
{Spatial-temporal redistribution of point defects in three-layer stressed
nanoheterosystems within the framework of self-assembled deformation-diffusion model}
\author[R.M.~Peleshchak, N.Ya.~Kulyk, M.V.~Doroshenko]%
{R.M.~Peleshchak, N.Ya.~Kulyk\thanks{E-mail:delenkonadia@mail.ru}\,, M.V.~Doroshenko}
\address{Drohobych Ivan Franko State Pedagogical University, 24~Franko St., 82100 Drohobych, Ukraine }

\date{Received November 20, 2014, in final form February 10, 2015}
\authorcopyright{R.M.~Peleshchak, N.Ya.~Kulyk, M.V.~Doroshenko, 2015}

\begin{document}

\maketitle

\begin{abstract}
The model of spatial-temporal distribution of point defects in a three-layer stressed nanoheterosystem \linebreak GaAs/In$_x$Ga$_{1 - x}$As/GaAs considering the self-assembled deformation-diffusion interaction is constructed.  Within  the framework of this model, the profile of spatial-temporal distribution of vacancies (interstitial atoms) in the stressed nanoheterosystem GaAs/In$_x$Ga$_{1 - x}$As/GaAs is calculated. It is shown that in the case of a stationary state ($t > 5\tau _d^{(2)}$), the concentration of vacancies in the inhomogeneously compressed interlayer is smaller relative to the initial average value $N_{d0}^{(2)}$ by 16\%.

\pacs  67.80.Mg, 68.55.Ln
\end{abstract}

\section{Introduction}

Intensive development of nanotechnologies has provided an opportunity to create nanoelectronic devices on the basis of stressed nanoheterosystems GaAs/In$_x$Ga$_{1 - x}$As/GaAs (ZnTe/Zn$_{1 - x}$Cd$_x$Te/ZnTe). The active region of such structures are layers In$_x$Ga$_{1 - x}$As, Zn$_{1 - x}$Cd$_x$Te, in which the electron-hole gas is localized being bounded on two sides of the potential barriers GaAs (ZnTe).

It is known that optical and electric properties of such devices depend significantly on both the lattice deformation of the contacting systems and the spatial distribution of point defects.

Such defects can penetrate from the surface or arise in the process of epitaxial growth. Besides, diffusion processes play an important role in the technology of fabricating optoelectronic devices. They are related with the redistribution of impurities in a semiconductor structure caused by both the ordinary gradient concentration of defects and the gradient of deformation tensor.

The interaction of defects with the deformation field, created by both the mismatch of the crystal lattice of the contacting materials and the point defects, causes a spatial redistribution of the latter. It can lead both to accumulation and to a decrease of the number of defects in the active region (In$_x$Ga$_{1 - x}$As, Zn$_{1 - x}$Cd$_x$Te) of the operating element depending on the character of the deformation created both by the mismatch between parameters of contacting crystal lattices (${\varepsilon _0} = 7\%$ ($4\%$) for GaAs/In$_x$Ga$_{1 - x}$As/GaAs  (ZnTe/Zn$_{1 - x}$Cd$_x$Te/ZnTe), respectively \cite{Led,Koz}), and by the action of defects. In particular, it is known that the gallium arsenide grown using the method of the molecular-beam epitaxy at low temperature contains an excess of arsenic \cite{Vil1,Vil2}. Introduction of excessive arsenic causes a tetragonal distortion of the lattice material GaAs and the generation of point defects in it: interstitial  atoms (As), vacancies (Ga) and antistructural defects (As$_\textrm{Ga}$), which, in turn, leads to their spatial redistribution. The lattice deformation and concentration of point defects generated under the action of the gradient of deformation tensor depend on the mismatch between the lattice parameters of contacting layers of heterostructure, the temperature of growth, molecular fluxes Ga and As, concentration and chemical nature of the doped impurities.

The strain caused by the mismatch between the lattice of the epitaxial layer and the substrate can be elastic when the thickness of the layer does not exceed a defined critical value \cite{Zhu}. Otherwise, mismatch dislocations are formed accompanied by a sharp worsening of both the optical and the electric characteristics of devices. However, in the layers In$_x$Ga$_{1 - x}$As with the mismatch less than critical there is a significant decline of the mobility and the intensity of photoluminescence at certain terms \cite{Zhu}, which is related to the increased number of point defects and a corresponding increase of the diffusion barrier to the atoms of the third group.

In experimental work \cite{Kar}, it is shown that in a heterostructure  GaAs/In$_x$Ga$_{1 - x}$As, the stressed \linebreak quantum-size heterolayers hamper the  diffusion of hydrogen and defects into the bulk of the material which leads to a substantial difference of their spatial distribution in a heterostructure and homogeneous layers. Theoretical research of the stationary distribution of defects within the framework of the self-assembled deformation-diffusion model has been considered in the work \cite{Pel}.

Therefore, in order to create devices with prescribed physical properties,  it is necessary to construct a spatial-temporal deformation-diffusion model that describes the self-assembled deformation-diffusion processes in stressed nanoheterostructures having their own point defects and impurities.

The aim of this work is to construct a spatial-temporal deformation-diffusion model and calculate the spatial-temporal profile distribution of point defects (interstitial atoms and vacancies) in three-layer stressed nanoheterosystems GaAs/In$_x$Ga$_{1 - x}$As/GaAs (ZnTe/Zn$_{1 - x}$Cd$_x$Te/ZnTe).

\section{The model of spatial-temporal redistribution of defects
in a three-layer stressed nanoheterosystem}

\begin{figure}[!b]
\centerline{
\includegraphics[width=0.67\textwidth]{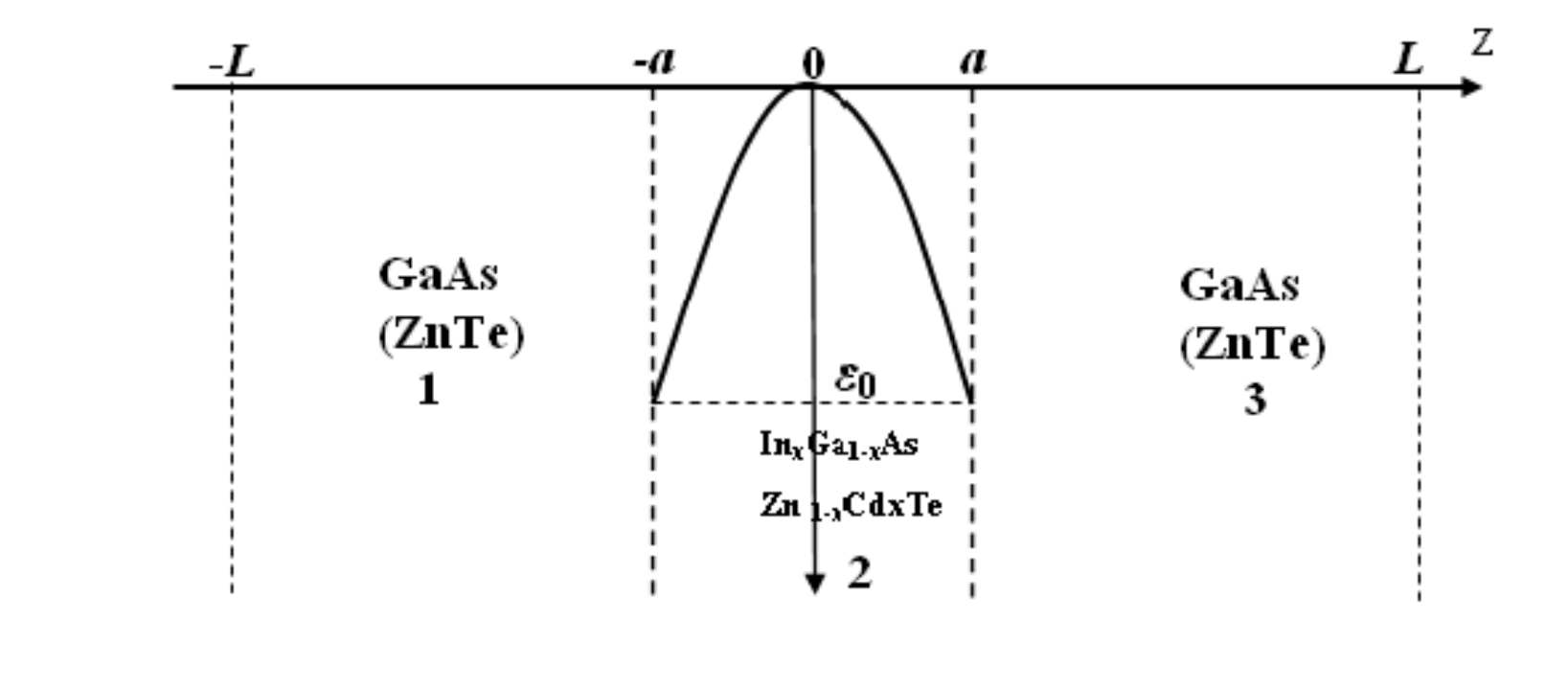}
}
\caption {The model of stressed nanoheterostructures
GaAs/In$_x$Ga$_{1 - x}$As/GaAs (ZnTe/Zn$_{1 - x}$Cd$_x$Te/ZnTe).} \label{fig1}
\end{figure}

Let us consider stressed nanoheterosystems GaAs/In$_x$Ga$_{1 - x}$As/GaAs  (ZnTe/Zn$_{1 - x}$Cd$_x$Te/ZnTe) having interlayers InAs (CdTe) of the thickness $2a$, that include three layers (figure~\ref{fig1}), where $N_{d0}^{(1)}$,  $N_{d0}^{(2)}$, $N_{d0}^{(3)}$ are the initial average defect concentrations, respectively, and ${D_1}$, ${D_2}$, ${D_3}$ are diffusion coefficients. Suppose that the external layers GaAs (ZnTe) are of the thickness that considerably exceeds the width of the interlayer of the heterostructure (${{2a}}/{L} \ll 1$), so deformation of these layers can be neglected [${\varepsilon _i}\left( z \right) = 0$, $i = 1,3$].

The mechanical deformation that occurs due to the mismatch between the lattice parameters of contacting materials of a heterosystem is approximated by the function \cite{Pel}:
\begin{equation}\label{1}
{\varepsilon _i}\left( z \right) = \left\{ {\begin{array}{*{20}{l}}
{{\varepsilon _0}\frac{{{z^2}}}{{{a^2}}},}&{i = 2,}\\
{0,}&{i = 1,3,}
\end{array}} \right.\
\end{equation}
where ${\varepsilon _0} = {\varepsilon _{xx}} + {\varepsilon _{yy}} + {\varepsilon _{zz}} < 0  $ is the relative change of the elementary cell volume  of the grown layer on heteroboundaries $z = \left| a \right|$; ${\varepsilon _{yy}} = {\varepsilon _{zz}} = {({a_{i + 1}} - {a_i})}/{{a_i}}$, ${\varepsilon _{xx}} =  - ({{2C_{_{12}}^{(2)}}}/{{C_{_{11}}^{(2)}}}){\varepsilon _{yy}}$, where $i = 1,3$ corresponds to the layers GaAs (ZnTe), $i = 2$ corresponds to the interlayers  In$_x$Ga$_{1 - x}$As (Zn$_{1 - x}$Cd$_x$Te); ${a_i}$ are the crystal lattice parameters of the contacting materials  GaAs (ZnTe) and InAs (CdTe) of the heterostructure, respectively; $C_{11}^{(2)}$ and $C_{12}^{(2)}$ are the elastic constants of the material In$_x$Ga$_{1 - x}$As (Zn$_{1 - x}$Cd$_x$Te).

Epitaxial growth on a substrate with the mismatch between the lattice parameters takes place simultaneously with the diffusion process, which is caused by both the concentration gradient of point defects [${\textrm{grad}}  N_{dl}^{(i)}(z,t)$] and the gradient parameter of deformation [${\textrm{grad}}  {U_i}\left( {z,t} \right)$]. The latter induces an additional diffusion flux of defects, which is opposite to the ordinary gradient concentration flux of defects. Therefore, in the basis of this model it is necessary to put a self-assembled system of non-stationary equations for the parameter deformation  ${U_i}\left( {z,t} \right)$ and concentration of impurities $N_d^{(i)}\left( {z,t} \right)$ in a stressed heterosystem, the redistribution of which is performed similarly to an ordinary diffusion flux,
\[J_\textrm{dif}^{(i)}\left( {z,t} \right)  =   -  {D_i}{{\frac{\partial N_d^{(i)}\left( {z,t} \right)}{\partial z}}},\]
thus, by the deformation component of the flux
\[J_\textrm{def}^{(i)}\left( {z,t} \right)  =   -  {D_i}{{{\theta _d^{(i)}} \over {{k_\textrm{B}}T}}}N_d^{(i)}\left( {z,t} \right){{\frac{\partial {U_i}\left( {z,t} \right)}{\partial z}}},\]
where $\theta _d^{\left( i \right)} = {K^{\left( i \right)}}\Delta {\Omega ^{\left( i \right)}}$ is the mechanical deformation potential, ${K^{(i)}}  = ({{{C_{11}^{(i)}  +  2C_{12}^{(i)}} }})/3$ is the module of uniform compression of the $i$-th material, $\Delta {\Omega ^{\left( i \right)}}$ is the variation of elementary cell volume at the presence of a defect in the $i$-th layer.

Let the point defects be distributed with the initial average concentration $N_{d0}^{(i)}$ in the $i$-th layer in a particular heterosystem. As a result of their self-assembled interaction through the deformation field, created by both the mismatch between lattice parameters of contacting materials of a heterosystem and the presence of defects, there is a variation of the concentration profile of point defects and of the character of deformation.

The mechanical stress in epitaxial layers created by both the point defects and the mismatch between lattice parameters of contacting materials is described by the expression:
\begin{equation}\label{2}
{\sigma _i}(z,t) = {\rho _i}c_i^2{U_i}(z,t) - \theta _d^{(i)}N_d^{(i)}(z,t) - {\rho _i}c_i^2\varepsilon {}_i(z),\
\end{equation}
where ${\rho _i}, {c_i}$ are the density of the $i$-th medium and the longitudinal speed of the sound, respectively.

The wave equation for the deformation parameter ${U_i}\left( {z,t} \right)$ is of the form:
\begin{equation}\label{3}
{\rho _i}\frac{{{\partial ^{^2}}{U_i}}}{{\partial {t^2}}} = \frac{{{\partial ^2}{\sigma _i}}}{{\partial {z^2}}}\,.
\end{equation}

Taking into account (\ref{2}), equation (\ref{3}) for the  renormalized deformation, ${U_i}\left( {z,t} \right)$ looks as follows:
\begin{equation}\label{4}
\frac{1}{{c_i^2}}\frac{{{\partial ^2}{U_i}\left( {z,t} \right)}}{{\partial {t^2}}} = \frac{{{\partial ^2}{U_i}(z,t)}}{{\partial {z^2}}} - \frac{{\theta _d^{(i)}}}{{{\rho _i}c_i^2}}\frac{{{\partial ^2}N_d^{(i)}(z,t)}}{{\partial {z^2}}} -  \frac{{{\partial ^2}{\varepsilon _i}(z)}}{{\partial {z^2}}}.\
\end{equation}

The equation for the defect concentration (interstitial atoms and vacancies) is of the form \cite{Pel}:
\begin{equation}\label{5}
\frac{{\partial N_d^{(i)}(z,t)}}{{\partial t}} = {D_i}\frac{{{\partial ^2}N_d^{(i)}(z,t)}}{{\partial {z^2}}}
- {D_i}\frac{{\theta _d^{(i)}}}{{{k_\textrm{B}}T}}\frac{\partial }{{\partial z}}\left[ {N_d^{(i)}\left( {z,t} \right)\frac{{\partial {U_i}(z,t)}}{{\partial z}}} \right] + G_d^{(i)} - \frac{{N_d^{(i)}(z,t)}}{{\tau _d^{(i)}}},\
\end{equation}
where ${D_i}$ is the diffusion coefficient of point defects in the $i$-th layer,
$ G_d^{(i)}$ is the generation rate of the defects, $\tau _d^{(i)}$ is the lifetime of the defects in the $i$-th layer that is determined by the frequency and the amplitude of mechanical fluctuations in the megahertz range ($\omega \geqslant 10^6$~Hz, $\tau _d^{(i)}\sim 1~{\mu}\textrm{s}$) that arise in the process of the formation of heteroboundaries in stressed nanoheterostructures and in the process of the occurrence of defects (acoustic emission) \cite{Pan}.

As a result, a self-assembled system of equations (\ref{4}), (\ref{5}) is received for determination of the spatial-temporal distribution of the concentration of defects $ N_d^{(i)}\left( {z,t} \right)$ and the deformation parameter ${U_i}\left( {z,t} \right)$ in the different regions of the three-layer nanoheterostructure.

The defect concentration can be written in the form:
\begin{equation}\label{6}
N_d^{(i)}(z,t) = N_{d0}^{(i)} + N_{dl}^{(i)}(z,t),\
\end{equation}
where $N_{dl}^{(i)}(z,t)$ is the spatially inhomogeneous component of the defect concentration.

Taking into account the presentation (\ref{6}) in the approximation $ N_{dl}^{(i)} \ll N_{d0}^{(i)}$, the equation of diffusion is written as follows:
\begin{equation}\label{7}
\frac{{\partial N_{dl}^{(i)}(z,t)}}{{\partial t}} = {D_i}\frac{{{\partial ^2}N_{dl}^{(i)}(z,t)}}{{\partial {z^2}}} - {D_i}N_{d0}^{(i)}\frac{{\theta _d^{(i)}}}{{{k_\textrm{B}}T}}\frac{{{\partial ^2}{U_i}\left( {z,t} \right)}}{{\partial {z^2}}} + G_d^{\prime(i)} - \frac{{N_{dl}^{(i)}(z,t)}}{{\tau _d^{(i)}}}\,,
\end{equation}
where $ G_d ^{\prime (i)} = G_d ^{(i)} - {{N_{d0}^{(i)}(z,t)}}/{{\tau _d^{(i)}}}$ is the generation rate of point defects under the effect of mechanical fluctuations ($\omega \sim 10^6 $~Hz) that arise in the process of the formation of heteroboundaries in stressed nanoheterostructures.

A further solution of self-assembled systems of equations (\ref{4}), (\ref{5}) will be searched in the approximation
\[{\left[ {{{\frac{\left({L_{d}^{(i)}}\right)^2}{2D_i^{}}}}} \right]^2}{{\frac{{\partial ^2}{U_i}\left( {z,t} \right)}{\partial {t^2}}}}  \ll {\varepsilon _0}, \] namely
\begin{equation}\label{8}
{\rm{ }}{{\frac{{\partial ^2}{U_i}\left( {z,t} \right)}{\partial {t^2}}}}  \approx 0,\
\end{equation}
where ${L_{di}}$ is the diffusion length of the defect in the $i$-th layer.

In approximation (\ref{8}), from equation (\ref{4}), there will be found ${\rm{ }}{{{{\partial ^2}{U_i}\left( {z,t} \right)} \over {\partial {z^2}}}} $  and it will be put into  the equation (\ref{7}). As a result, differential equation for determination of the spatial-temporal distribution of defects in the stressed heterosystem is received
\begin{equation}\label{9}
\frac{\partial {N_{dl}^{(i)}(z,t)}}{{\partial t}} = \left[ {{D_i}\left( {1  - {{{N_{d0}^{(i)}} \over {N_{dc}^{(i)}}}}} \right)} \right]\frac{{{\partial ^2}N_{dl}^{(i)}(z,t)}}{{\partial {z^2}}} -  {D_i} {{{N_{d0}^{(i)}} \over {N_{dc}^{(i)}\Delta \Omega }}}\frac{{{\partial ^2}{\varepsilon _i}(z)}}{{\partial {z^2}}}  + G_d^{\prime(i)}  -  \frac{{N_{dl}^{(i)}(z,t)}}{{\tau _d^{(i)}}} ,\
\end{equation}
where $ N_{dc}^{(i)}  =  {{{{k_\textrm{B}}T \rho  c_i^2} / {\theta _d^{(i)}}}}$ is the critical defect concentration, which being exceeded results in the self-organization of the defects \cite{Eme}.

In addition, the conditions of the equality of concentration of impurities and their fluxes must be satisfied on the boundary layer of the heterostructure shown in figure~\ref{fig1}:
%
%
\[\frac{{\partial {N_{dl}^{\left( 1 \right)}}(z,t)}}{{\partial z }}{\Bigg|_{z \to -\infty}} = 0,       \qquad                     \frac{{\partial {N_{dl}^{\left( 3 \right)}}(z,t)}}{{\partial z }}{\Bigg|_{z \to \infty}} = 0,\]
\begin{equation}\label{10}
N_{dl}^{\left( 1 \right)}( - a,t) = N_{dl}^{\left( 2 \right)}( - a,t),  \qquad  N_{dl}^{\left( 2 \right)}(a,t) = N_{dl}^{\left( 3 \right)}(a,t),\
\end{equation}
\[{J_1}\left( { - a,t} \right)  =  {J_2}\left( { - a,t}\right), \qquad
{J_2}\left( {a,t} \right)  =  {J_3}\left( {a,t} \right),
        \]
where ${J_i}\left( {z,t} \right) =  - \frac{\partial }{{\partial z}}\left\{ {{D_i}N_{dl}^{(i)}(z,t)[1 - {\varepsilon _i}(z)]} \right\}$.
At the primary moment of time
\begin{equation}
N_{dl}^{(i)}\left( {z,0} \right)  =  0.\
\end{equation}

Entering the following dimensionless variables
\[\theta   =  {{t \over {\tau _d^{\left( 2 \right)}}}}, \qquad L_d^{\left( i \right)}  = \sqrt {{D_i}\tau _d^{\left( i \right)}} ,  \]
\[N_{dl}^{\left( i \right)}\left( {z,t} \right)  =  {Y_i}\left( {z,t} \right)G_d^{\prime\left( i \right)}\tau _d^{\left( i \right)},\]
\[   \mathord{\buildrel{\lower3pt\hbox{$\scriptscriptstyle\frown$}}
\over z}   =  {{z \over L}},\qquad \mathord{\buildrel{\lower3pt\hbox{$\scriptscriptstyle\frown$}}
\over L}   =  {{L \over a}},
\qquad
N_{d0}^{\left( i \right)}  =  G_d^{\prime\left( i \right)}\tau _d^{\left( i \right)}, \]
\begin{equation}\label{12}
\beta   =   {D_2}{{{N_{d0}^{\left( 2 \right)}} \over {N_{dc}^{\left( 2 \right)}}}}{{{2{\varepsilon _0}} \over {\Delta {\Omega ^{\left( 2 \right)}} {a^2}G_d^{\left( 2 \right)}}}}\,,
\end{equation}
equations (\ref{9}) take the form:
\begin{align}
&\frac{{\partial {Y_1}(\mathord{\buildrel{\lower3pt\hbox{$\scriptscriptstyle\frown$}}
\over z} ,\theta )}}{{\partial \theta }} = {{{{D_1}} \over {{D_2}}}}{\left( {{{{L_d^{\left( 2 \right)}} \over {{L}}}}} \right)^2}\left( {1  - {{{N_{d0}^{(1)}} \over {N_{dc}^{(1)}}}}} \right)\frac{{{\partial ^2}{Y_1}(\mathord{\buildrel{\lower3pt\hbox{$\scriptscriptstyle\frown$}}
\over z} ,\theta )}}{{\partial {{\mathord{\buildrel{\lower3pt\hbox{$\scriptscriptstyle\frown$}}
\over z} }^2}}} - \lambda_1\left[ {{Y_1}(\mathord{\buildrel{\lower3pt\hbox{$\scriptscriptstyle\frown$}}
\over z} ,\theta)  - 1} \right],\nonumber\\
%
\label{13}
&\frac{{\partial {Y_2}(\mathord{\buildrel{\lower3pt\hbox{$\scriptscriptstyle\frown$}}
\over z} ,\theta )}}{{\partial \theta }} = {\left( {{{{L_d^{\left( 2 \right)}} \over L}}} \right)^2}\left( {1  -  {{{N_{d0}^{(2)}} \over {N_{dc}^{(2)}}}}} \right)\frac{{{\partial ^2}{Y_2}(\mathord{\buildrel{\lower3pt\hbox{$\scriptscriptstyle\frown$}}
\over z} ,\theta )}}{{\partial {{\mathord{\buildrel{\lower3pt\hbox{$\scriptscriptstyle\frown$}}
\over z} }^2}}}  +  \beta  -\lambda_2 \left[ {{Y_2}(\mathord{\buildrel{\lower3pt\hbox{$\scriptscriptstyle\frown$}}
\over z} ,\theta)  - 1} \right],\\
%
&\frac{{\partial {Y_3}(\mathord{\buildrel{\lower3pt\hbox{$\scriptscriptstyle\frown$}}
\over z} ,\theta )}}{{\partial \theta }} = {{{{D_3}} \over {{D_2}}}}{\left( {{{{L_d^{\left( 2 \right)}} \over {{L}}}}} \right)^2}\left( {1  - {{{N_{d0}^{(3)}} \over {N_{dc}^{(3)}}}}} \right)\frac{{{\partial ^2}{Y_3}(\mathord{\buildrel{\lower3pt\hbox{$\scriptscriptstyle\frown$}}
\over z} ,\theta )}}{{\partial {{\mathord{\buildrel{\lower3pt\hbox{$\scriptscriptstyle\frown$}}
\over z} }^2}}} - \lambda_3\left[ {{Y_3}(\mathord{\buildrel{\lower3pt\hbox{$\scriptscriptstyle\frown$}}
\over z} ,\theta)  - 1} \right],\nonumber
\end{align}
where $ \lambda_1 = {{{{D_1}} \over {{D_2}}}}{\left( {{{{L_d^{\left( 2 \right)}} / {{L_d^{\left( 1 \right)}}}}}} \right)^2}$, $ \lambda_2 =1$, $ \lambda_3 = {{{{D_3}} \over {{D_2}}}}{\left( {{{{L_d^{\left( 2 \right)}} / {{L_d^{\left( 3 \right)}}}}}} \right)^2}$,
and boundary conditions (\ref{10}) can be written:
%
\[\frac{{\partial {Y_1}(\mathord{\buildrel{\lower3pt\hbox{$\scriptscriptstyle\frown$}}
\over z} ,\theta )}}{{\partial \mathord{\buildrel{\lower3pt\hbox{$\scriptscriptstyle\frown$}}
\over z} }}{\bigg|_{\mathord{\buildrel{\lower3pt\hbox{$\scriptscriptstyle\frown$}}
\over z} \to -\mathord{\buildrel{\lower3pt\hbox{$\scriptscriptstyle\frown$}}
\over L} }} = 0,\]
\[ {Y_1}( - \mathord{\buildrel{\lower3pt\hbox{$\scriptscriptstyle\frown$}}
\over a} ,\theta ) = {{{G_d^{\left( 2 \right)} \tau _d^{\left( 2 \right)}} \over {G_d^{\left( 1 \right)}\tau _d^{\left( 1 \right)} }}}{Y_2}( - \mathord{\buildrel{\lower3pt\hbox{$\scriptscriptstyle\frown$}}
\over a} ,\theta ),\]
\begin{equation}\label{14}
{Y_2}(\mathord{\buildrel{\lower3pt\hbox{$\scriptscriptstyle\frown$}}
\over a} ,\theta ) = {{{G_d^{\left( 3 \right)} \tau _d^{\left( 3 \right)}} \over {G_d^{\left( 2 \right)}\tau _d^{\left( 2 \right)} }}}{Y_3}(\mathord{\buildrel{\lower3pt\hbox{$\scriptscriptstyle\frown$}}
\over a} ,\theta ),\
\end{equation}
\[{{{{D_1}} \over {{D_2}}}}{\left( {{{{L_d^{\left( 2 \right)}} \over {{L}}}}} \right)^2}\left( { - 1  +  {{{N_{d0}^{\left( 1 \right)}} \over {N_{dc}^{\left( 1 \right)}}}}} \right){{{\partial {Y_1}( { - \mathord{\buildrel{\lower3pt\hbox{$\scriptscriptstyle\frown$}}
\over a} , \theta} )} \over {\partial \mathord{\buildrel{\lower3pt\hbox{$\scriptscriptstyle\frown$}}
\over z} }}}  = {\left( {{{{L_d^{\left( 2 \right)}} \over L}}} \right)^2} \left( { - 1  +  {{{N_{d0}^{\left( 2 \right)}} \over {N_{dc}^{\left( 2 \right)}}}}} \right){{{\partial {Y_2}( { - \mathord{\buildrel{\lower3pt\hbox{$\scriptscriptstyle\frown$}}
\over a} , \theta} )} \over {\partial \mathord{\buildrel{\lower3pt\hbox{$\scriptscriptstyle\frown$}}
\over z} }}}   -  \beta \mathord{\buildrel{\lower3pt\hbox{$\scriptscriptstyle\frown$}}
\over a} ,\]
\[{\left( {{{{L_d^{\left( 2 \right)}} \over L}}} \right)^2} \left( { - 1  + {{{N_{d0}^{\left( 2 \right)}} \over {N_{dc}^{\left( 2 \right)}}}}} \right){{{\partial {Y_2}( { \mathord{\buildrel{\lower3pt\hbox{$\scriptscriptstyle\frown$}}
\over a} , \theta} )} \over {\partial \mathord{\buildrel{\lower3pt\hbox{$\scriptscriptstyle\frown$}}
\over z} }}}   +  \beta \mathord{\buildrel{\lower3pt\hbox{$\scriptscriptstyle\frown$}}
\over a}    = {{{{D_3}} \over {{D_2}}}}{\left( {{{{L_d^{\left( 2 \right)}} \over {{L}}}}} \right)^2}  \left( { - 1  +  {{{N_{d0}^{\left( 3 \right)}} \over {N_{dc}^{\left( 3 \right)}}}}} \right){{{\partial {Y_3}( {\mathord{\buildrel{\lower3pt\hbox{$\scriptscriptstyle\frown$}}
\over a} , \theta} )} \over {\partial \mathord{\buildrel{\lower3pt\hbox{$\scriptscriptstyle\frown$}}
\over z} }}},\]
\[\frac{{\partial {Y_3}(\mathord{\buildrel{\lower3pt\hbox{$\scriptscriptstyle\frown$}}
\over z} ,\theta )}}{{\partial \mathord{\buildrel{\lower3pt\hbox{$\scriptscriptstyle\frown$}}
\over z} }}{\bigg|_{\mathord{\buildrel{\lower3pt\hbox{$\scriptscriptstyle\frown$}}
\over z} \to \mathord{\buildrel{\lower3pt\hbox{$\scriptscriptstyle\frown$}}
\over L} }} = 0.
\]

As seen from equation (\ref{12}), parameter $\beta$ describes the nature of the deformation effect caused both by the action of the stressed heteroboundary and the action of point defects of the type of compression or tension centers.
This parameter can take both the positive values  $\beta   >  0$ (${\varepsilon _0}  > 0$, $\Delta {\Omega ^{(2)}}  >  0$; ${\varepsilon _0}  <  0$, $\Delta {\Omega ^{(2)}}  <  0$) and the negative values $\beta   <  0$ (${\varepsilon _0}  >  0$, $\Delta {\Omega ^{(2)}}  < 0$; ${\varepsilon _0}  <  0$, $\Delta {\Omega ^{(2)}} >  0$).

The solution of equations (\ref{13}) with boundary conditions (\ref{14}) is searched in the form:
\begin{equation}\label{15}
{Y_i}( {\mathord{\buildrel{\lower3pt\hbox{$\scriptscriptstyle\frown$}}
\over z} , \theta } )  =  {\re^{ - \lambda_i \theta }} {Z_i}( {\mathord{\buildrel{\lower3pt\hbox{$\scriptscriptstyle\frown$}}
\over z} , \theta } ),
\end{equation}
where $ {Z_i}( {\mathord{\buildrel{\lower3pt\hbox{$\scriptscriptstyle\frown$}}
\over z} , \theta } )$  satisfy the following equations:
%
\begin{align}
&\frac{{\partial {Z_1}(\mathord{\buildrel{\lower3pt\hbox{$\scriptscriptstyle\frown$}}
\over z} ,\theta )}}{{\partial \theta }} = {{{{D_1}} \over {{D_2}}}}{\left( {{{{L_d^{\left( 2 \right)}} \over {{L}}}}} \right)^2}\left( {1  - {{{N_{d0}^{(1)}} \over {N_{dc}^{1}}}}} \right)\frac{{{\partial ^2}{Z_1}(\mathord{\buildrel{\lower3pt\hbox{$\scriptscriptstyle\frown$}}
\over z} ,\theta )}}{{\partial {{\mathord{\buildrel{\lower3pt\hbox{$\scriptscriptstyle\frown$}}
\over z} }^2}}}  +  \lambda_1{\re^{\lambda_1 \theta}},\nonumber\\
%
\label{16}
&\frac{{\partial {Z_2}(\mathord{\buildrel{\lower3pt\hbox{$\scriptscriptstyle\frown$}}
\over z} ,\theta )}}{{\partial \theta }} = {\left( {{{{L_d^{\left( 2 \right)}} \over L}}} \right)^2}\left( {1  -  {{{N_{d0}^{(2)}} \over {N_{dc}^{(2)}}}}} \right)\frac{{{\partial ^2}{Z_2}(\mathord{\buildrel{\lower3pt\hbox{$\scriptscriptstyle\frown$}}
\over z} ,\theta )}}{{\partial {{\mathord{\buildrel{\lower3pt\hbox{$\scriptscriptstyle\frown$}}
\over z} }^2}}}  +  {\re^\theta }\left( {\beta   +  1} \right),\\
%
&\frac{{\partial {Z_3}(\mathord{\buildrel{\lower3pt\hbox{$\scriptscriptstyle\frown$}}
\over z} ,\theta )}}{{\partial \theta }} ={{{{D_3}} \over {{D_2}}}}{\left( {{{{L_d^{\left( 2 \right)}} \over {{L}}}}} \right)^2} \left( {1  - {{{N_{d0}^{(3)}} \over {N_{dc}^{(3)}}}}} \right)\frac{{{\partial ^2}{Z_3}(\mathord{\buildrel{\lower3pt\hbox{$\scriptscriptstyle\frown$}}
\over z} ,\theta )}}{{\partial {{\mathord{\buildrel{\lower3pt\hbox{$\scriptscriptstyle\frown$}}
\over z} }^2}}}  +  \lambda_3{\re^{\lambda_3 \theta}},\nonumber
\end{align}
with boundary conditions:
\[\frac{{\partial {Z_1}(\mathord{\buildrel{\lower3pt\hbox{$\scriptscriptstyle\frown$}}
\over z} ,\theta )}}{{\partial \mathord{\buildrel{\lower3pt\hbox{$\scriptscriptstyle\frown$}}
\over z} }}{\bigg|_ {\mathord{\buildrel{\lower3pt\hbox{$\scriptscriptstyle\frown$}}
\over z} \to -\mathord{\buildrel{\lower3pt\hbox{$\scriptscriptstyle\frown$}}
\over L} }} = 0, \]
\[
{\re^{-\lambda_1 \theta}} {Z_1}( - \mathord{\buildrel{\lower3pt\hbox{$\scriptscriptstyle\frown$}}
\over a} ,\theta ) = {{{G_d^{\left( 2 \right)} \tau _d^{\left( 2 \right)}} \over {G_d^{\left( 1 \right)}\tau _d^{\left( 1 \right)} }}}{\re^{-\theta}} {Z_2}( - \mathord{\buildrel{\lower3pt\hbox{$\scriptscriptstyle\frown$}}
\over a} ,\theta ),         \
\]
\begin{equation}\label{17}
{\re^{- \theta}}{Z_2}(\mathord{\buildrel{\lower3pt\hbox{$\scriptscriptstyle\frown$}}
\over a} ,\theta ) = {{{G_d^{\left( 3 \right)} \tau _d^{\left( 3 \right)}} \over {G_d^{\left( 2 \right)}\tau _d^{\left( 2 \right)} }}}{\re^{-\lambda_3 \theta}} {Z_3}(\mathord{\buildrel{\lower3pt\hbox{$\scriptscriptstyle\frown$}}
\over a} ,\theta ),
\end{equation}
\[{{{{D_1}} \over {{D_2}}}}{\left( {{{{L_d^{\left( 2 \right)}} \over {{L}}}}} \right)^2}{\re^{-\lambda_1 \theta}}\left( { - 1  +  {{{N_{d0}^{\left( 1 \right)}} \over {N_{dc}^{\left( 1 \right)}}}}} \right){{{\partial {Z_1}( { - \mathord{\buildrel{\lower3pt\hbox{$\scriptscriptstyle\frown$}}
\over a} , \theta} )} \over {\partial \mathord{\buildrel{\lower3pt\hbox{$\scriptscriptstyle\frown$}}
\over z} }}}  ={\left( {{{{L_d^{\left( 2 \right)}} \over L}}} \right)^2}{\re^{- \theta}}  \left( { - 1  +  {{{N_{d0}^{\left( 2 \right)}} \over {N_{dc}^{\left( 2 \right)}}}}} \right){{{\partial {Z_2}( { - \mathord{\buildrel{\lower3pt\hbox{$\scriptscriptstyle\frown$}}
\over a} , \theta} )} \over {\partial \mathord{\buildrel{\lower3pt\hbox{$\scriptscriptstyle\frown$}}
\over z} }}}   - \beta \mathord{\buildrel{\lower3pt\hbox{$\scriptscriptstyle\frown$}}
\over a} ,\]
\[{\left( {{{{L_d^{\left( 2 \right)}} \over L}}} \right)^2}{\re^{-\theta }} \left( { - 1  +  {{{N_{d0}^{\left( 2 \right)}} \over {N_{dc}^{\left( 2 \right)}}}}} \right){{{\partial {Z_2}( { \mathord{\buildrel{\lower3pt\hbox{$\scriptscriptstyle\frown$}}
\over a} , t} )} \over {\partial \mathord{\buildrel{\lower3pt\hbox{$\scriptscriptstyle\frown$}}
\over z} }}}   + \beta \mathord{\buildrel{\lower3pt\hbox{$\scriptscriptstyle\frown$}}
\over a}  = {{{{D_3}} \over {{D_2}}}}{\left( {{{{L_d^{\left( 2 \right)}} \over {{L}}}}} \right)^2}{\re^{-\lambda_3 \theta}}  \left( { - 1  + {{{N_{d0}^{\left( 3 \right)}} \over {N_{dc}^{\left( 3 \right)}}}}} \right){{{\partial {Z_3}( {\mathord{\buildrel{\lower3pt\hbox{$\scriptscriptstyle\frown$}}
\over a} , t} )} \over {\partial \mathord{\buildrel{\lower3pt\hbox{$\scriptscriptstyle\frown$}}
\over z} }}},\]
\[\frac{{\partial {Z_3}(\mathord{\buildrel{\lower3pt\hbox{$\scriptscriptstyle\frown$}}
\over z} ,\theta )}}{{\partial \mathord{\buildrel{\lower3pt\hbox{$\scriptscriptstyle\frown$}}
\over z} }}{\bigg|_ {\mathord{\buildrel{\lower3pt\hbox{$\scriptscriptstyle\frown$}}
\over z} \to \mathord{\buildrel{\lower3pt\hbox{$\scriptscriptstyle\frown$}}
\over L} }} = 0.\]

Solutions of equations (\ref{16}) with boundary conditions (\ref{17}) are presented in the appendix at
\[ \lambda_1= \lambda_2= \lambda_3=1,
\qquad {{{G_d^{\left( 2 \right)} \tau _d^{\left( 2 \right)}} \over {G_d^{\left( 1 \right)}\tau _d^{\left( 1 \right)} }}}=1,
\qquad
{{{G_d^{\left( 2 \right)} \tau _d^{\left( 2 \right)}} \over {G_d^{\left( 3 \right)}\tau _d^{\left( 3 \right)} }}}=1,
\qquad
{{{{D_1}} \over {{D_2}}}}{\left( {{{{L_d^{\left( 2 \right)}} \over {{a}}}}} \right)^2}=1, \qquad
{{{{D_3}} \over {{D_2}}}}{\left( {{{{L_d^{\left( 2 \right)}} \over {{a}}}}} \right)^2}=1.
\]

\section{Analysis of the numerical results and discussion}

\begin{figure}[!b]
\centerline{
\includegraphics[width=0.65\textwidth]{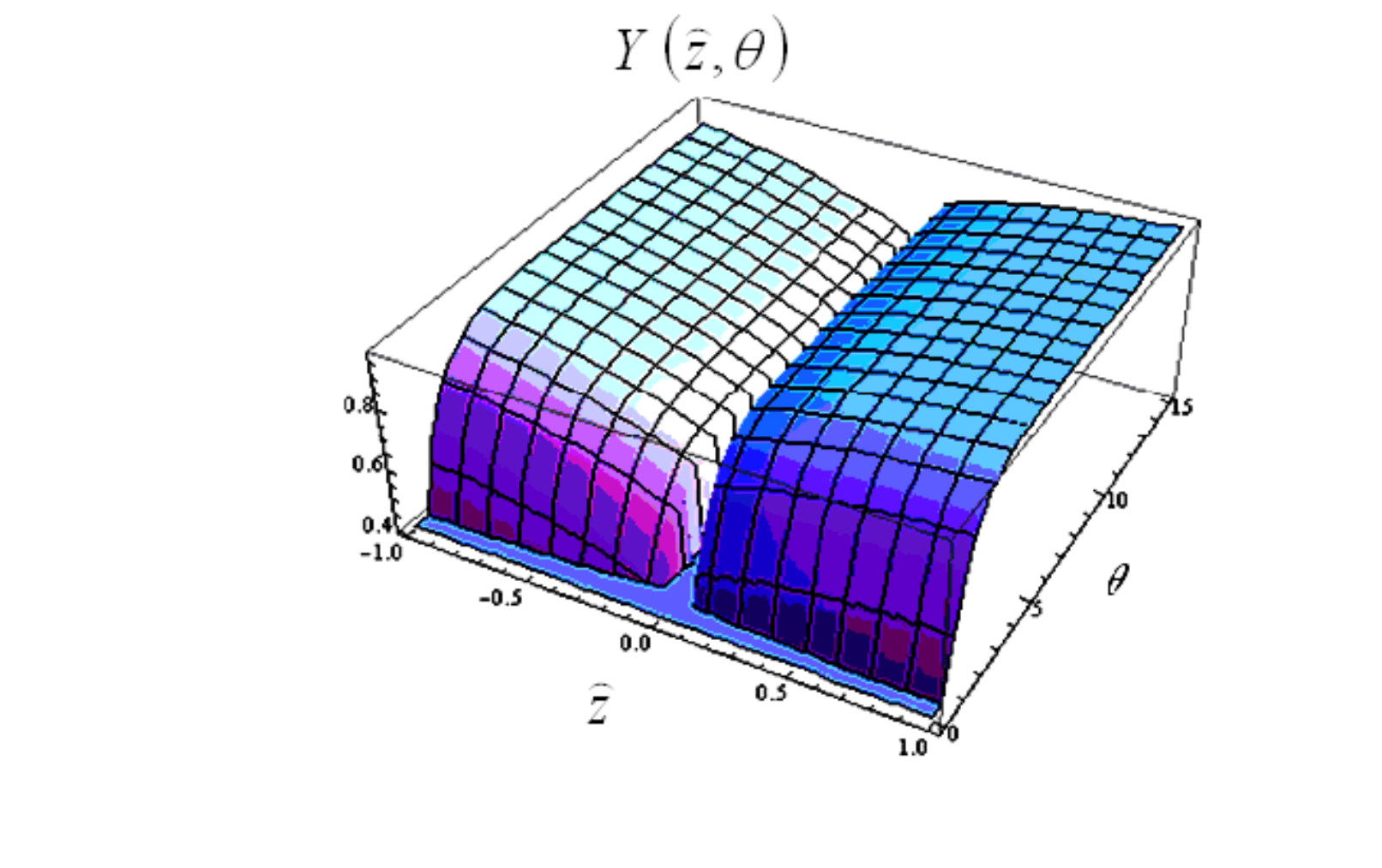}
}
\vspace{-5mm}
\caption{(Color online) Profile of the spatial-temporal distribution of the vacancies concentration in a three-layer stressed nanoheterosystem having inhomogeneous-compressed interlayer ${{{N_{d0}^{\left( 2 \right)}} / {N_{dc}^{\left( 2 \right)}}}}  =  0.5$; ${{{N_{d0}^{\left( 2 \right)}} / {N_{dc}^{\left( 2 \right)}}}}  =  0.8$; ${{{N_{d0}^{\left( 2 \right)}} / {N_{dc}^{\left( 2 \right)}}}}  =  0.6$; $\beta = 10.2$ [formula (\ref{12})].} \label{fig2}
\end{figure}

\begin{figure}[!t]
\centerline{
\includegraphics[width=0.65\textwidth]{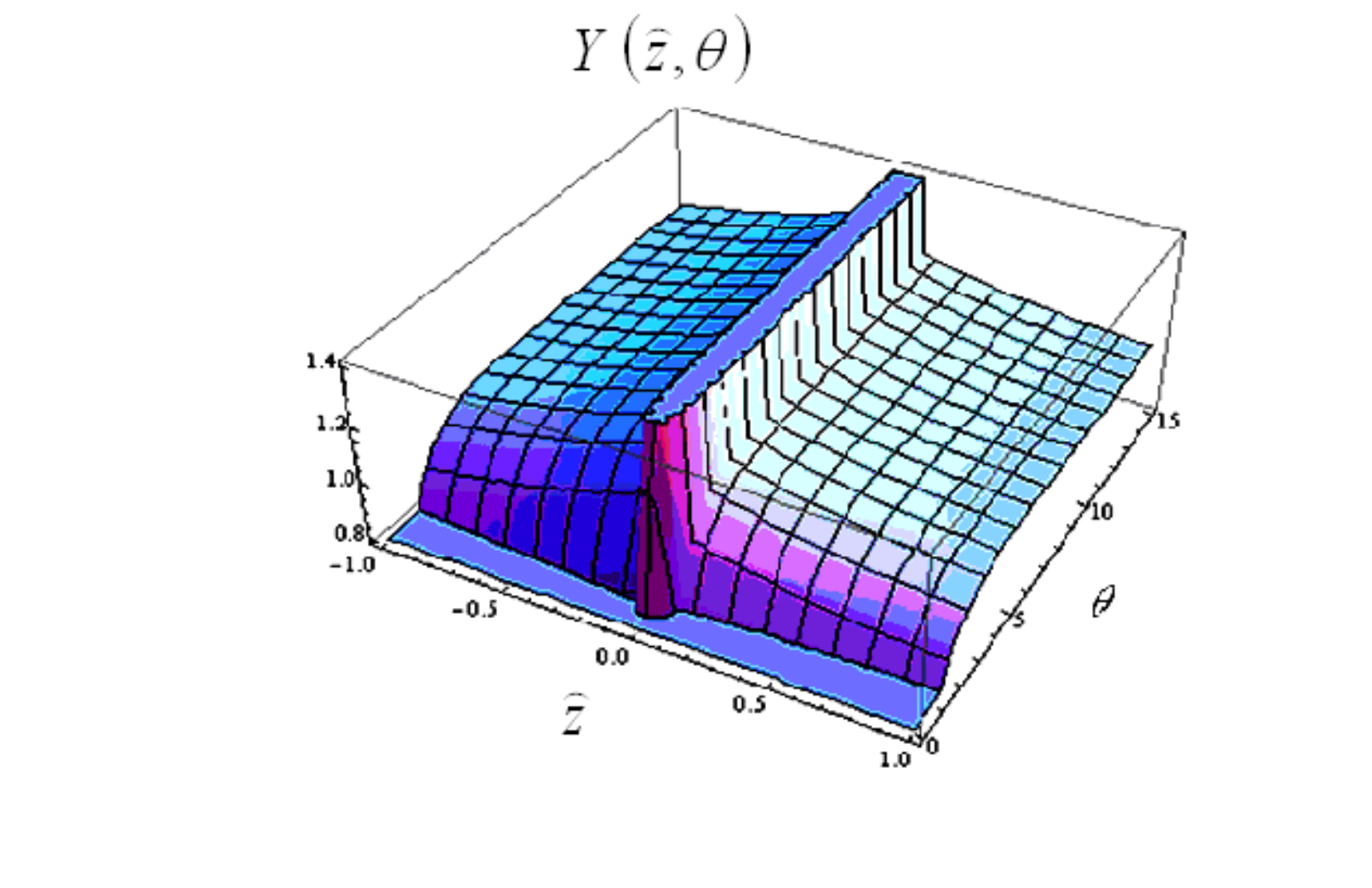}
}
\vspace{-5mm}
\caption{(Color online) Profile of the spatial-temporal distribution of the concentration of interstitial atoms in a three-layer stressed nanoheterosystem having inhomogeneous-compressed interlayer ${{{N_{d0}^{\left( 2 \right)}} / {N_{dc}^{\left( 2 \right)}}}}  = 0.5$; ${{{N_{d0}^{\left( 2 \right)}} / {N_{dc}^{\left( 2 \right)}}}}  = 0.8$; ${{{N_{d0}^{\left( 2 \right)}} / {N_{dc}^{\left( 2 \right)}}}}  =  0.6$; $\beta = -10.2$ [formula (\ref{12})].}  \label{fig3}
\end{figure}

In figures~\ref{fig2} and \ref{fig3}) there is shown the spatial-temporal redistribution of vacancies ${Y}({\mathord{\buildrel{\lower3pt\hbox{$\scriptscriptstyle\frown$}}
\over z} , \theta } )$ (figure~\ref{fig2}) and interstitial atoms (figure~\ref{fig3}) in a three-layer stressed nanoheterosystem GaAs/In$_x$Ga$_{1 - x}$As/GaAs under the effect of the deformation caused by both the mismatch between parameters of the contacting lattices (${\varepsilon _0}={{\Delta{a}}}/{a}=7\%$) and by the action of a point defect. Calculations were carried out for the following values of the parameters: ${\varepsilon _0}=0.07$; $a=0.05L$, $0.1L$ ($L$~--- the thickness of nanoheterostructure); ${{{N_{d0}^{\left( 1 \right)}} / {N_{dc}^{\left( 1 \right)}}}}  = 0.5$; ${{{N_{d0}^{\left( 2 \right)}} / {N_{dc}^{\left( 2 \right)}}}}  = 0.8$; ${{{N_{d0}^{\left( 3 \right)}} / {N_{dc}^{\left( 3 \right)}}}}  = 0.6$; $C_{11}^{(2)}=0.833$~Mbar; $C_{12}^{(2)}=0.453$~Mbar; $C_{11}^{(1)}=1.223$~Mbar; $C_{12}^{(1)}=0.571$~Mbar; $T=300$~K; $\theta _d^{\left( i \right)}=5$~eV \cite{Eme}; ${D_{i}}=10^{-5}$~{cm$^{2}$}/{s}; $\tau _d^{\left( 2 \right)}=1$~\textmu{}{s}.

As shown in figures~\ref{fig2} and \ref{fig3}, the profile of the spatial-temporal distribution of the defect concentration of the type of compression (vacancies, figure~\ref{fig2}) or tension (figure~\ref{fig3}) centers in a three-layer stressed nanoheterosystem is of a nonmonotonous character. If an internal epitaxial layer undergoes an inhomogeneous compression deformation (${\varepsilon _0}  <  0$) due to the mismatch between the lattice parameter of the contacting epitaxial layers, then a decrease (an increase) of the concentration of vacancies (interstitial atoms) in the interlayer of the three-layer nanoheterostructures will be observed.

If the epitaxial layer undergoes the tension deformation due to a mismatch between the lattice parameter of the epitaxial layer and the substrate (${a _\textrm{s}}  > {a_0}$, ${\varepsilon _0}  >  0$, where ${a _\textrm{s}}$ is the lattice parameter of the substrate; ${a _0}$ is the lattice parameter of the stackable layer), the opposite effect will be observed: near the heteroboundary there will be accumulation of vacancies and a decrease of the concentration of interstitial atoms. This, in turn, will lead to a decrease of the tension deformation in the epitaxial layer near the heteroboundary.

The effect of impoverishment (enrichment) in the interlayer of vacancies (interstitial atoms) has been observed in experimental works \cite{Kar,Che} after the growth (decline) of the intensity of photoluminescence in stressed nanoheterostructures.

\begin{figure}[!b]
\centerline{
\includegraphics[width=0.65\textwidth]{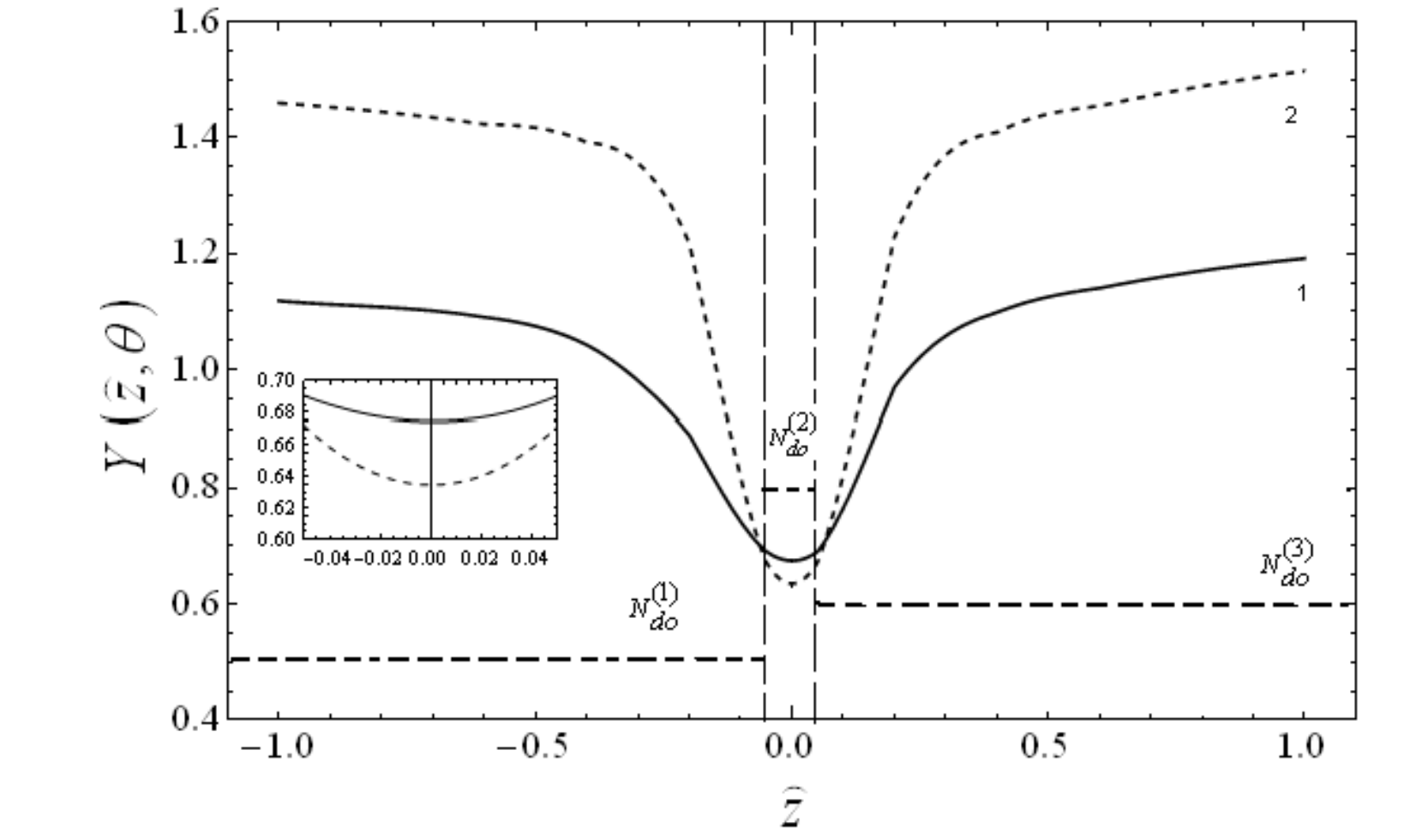}
}
\caption{The cut of the spatial-temporal distribution of the  concentration of vacancies along the growth axis  at different times: 1~--- at the moment $ t=\tau _d^{\left( 2 \right)}$; 2~--- $ t= 5\tau _d^{\left( 2 \right)}$; ${{{N_{d0}^{\left( 1 \right)}} / {N_{dc}^{\left( 1 \right)}}}}  = 0.5$; ${{{N_{d0}^{\left( 2 \right)}} / {N_{dc}^{\left( 2 \right)}}}}  = 0.8$; ${{{N_{d0}^{\left( 3 \right)}} / {N_{dc}^{\left( 3 \right)}}}}  = 0.6$; $\beta =10.2$ [formula (\ref{12})]; $ a=0.05L$.} \label{fig4}
\end{figure}

\begin{figure}[!t]
\centerline{
\includegraphics[width=0.65\textwidth]{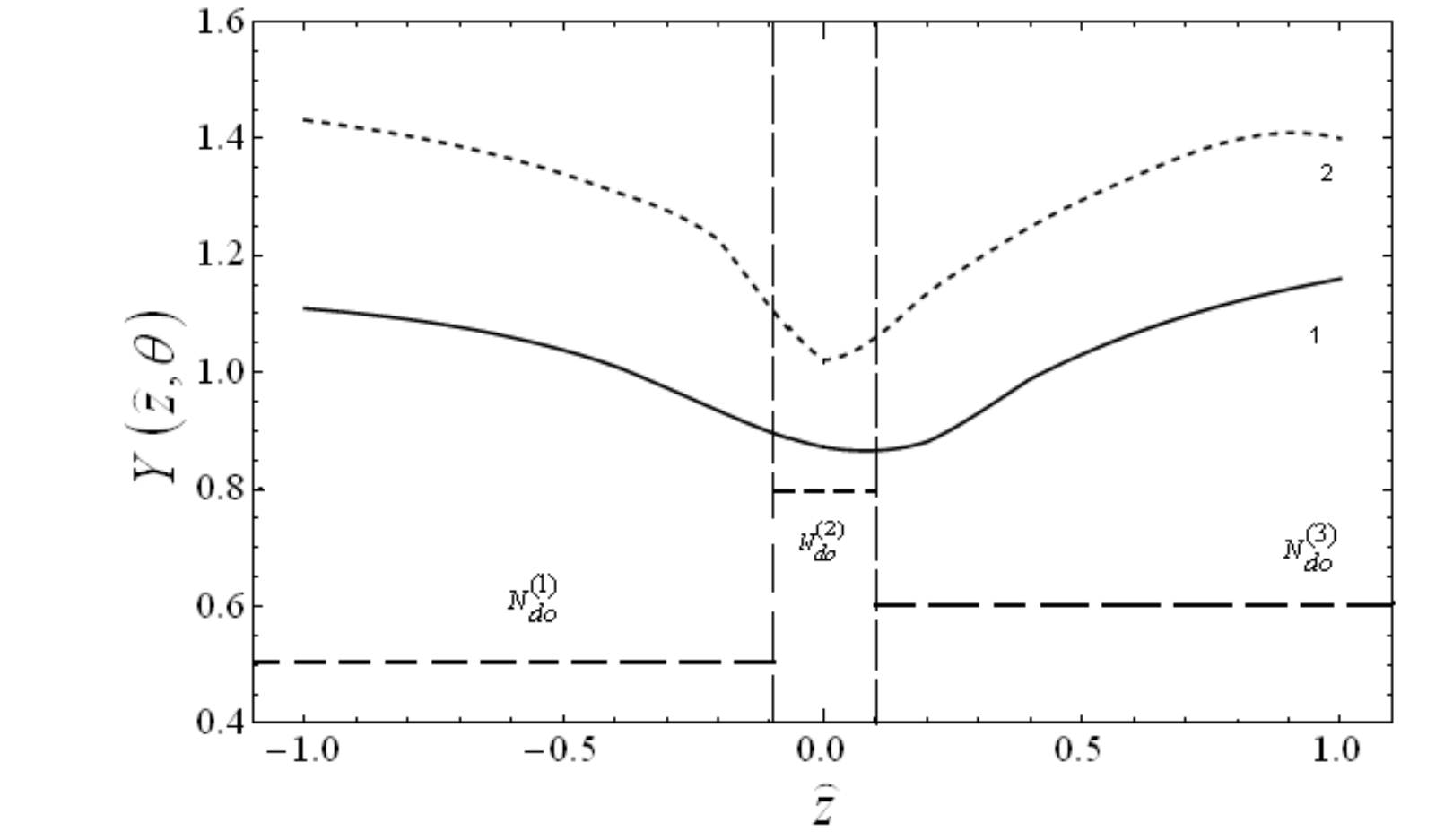}
}
\caption{The cut of the spatial-temporal distribution of the concentration of interstitial atoms along the growth axis  at different times: 1~--- at the moment $ t=\tau _d^{\left( 2 \right)}$; 2~--- $ t= 5\tau _d^{\left( 2 \right)}$; ${{{N_{d0}^{\left( 1 \right)}} / {N_{dc}^{\left( 1 \right)}}}}  = 0.5$; ${{{N_{d0}^{\left( 2 \right)}} / {N_{dc}^{\left( 2 \right)}}}}  = 0.8$; ${{{N_{d0}^{\left( 3 \right)}} / {N_{dc}^{\left( 3 \right)}}}}  = 0.6$; $\beta =10.2$ [formula (\ref{12})]; $a=0.1L$.} \label{fig5}
\end{figure}

In figures~\ref{fig4} and \ref{fig5}, numerical calculations of the cut of the spatial-temporal distribution of the  concentration of vacancies along the growth axis ($OX$) of heterosystem at different times are presented: $t=0$; $ \tau _d^{\left( 2 \right)}$; $ 5\tau _d^{\left( 2 \right)}$ ($ \tau _d^{\left( 2 \right)}$ is the average time of finding the defect in one of the equilibrium positions in the interlayer nanoheterosystem, namely a settled life) and for the different thicknesses of the interlayer of a nanoheterostructure ($a=0.05L$, figure~\ref{fig4} and $a=0.1L$, figure~\ref{fig5}).

As seen from figures~\ref{fig4} and \ref{fig5}, during the time interval $ 0  \leqslant  t   \leqslant 5 \tau _d^{\left( 2 \right)}$ there occurs a spatial-temporal redistribution of the defects, so that in the inhomogeneous-compressed interlayer [figure~\ref{fig1}, region~2, formula (\ref{1})] they become smaller relative to their initial average value $ N{}_{d0}^{\left( 2 \right)}$  by $ \approx 13.7 \% ,16 \% $ at different times $ \tau _d^{\left( 2 \right)}$, $ 5\tau _d^{\left( 2 \right)}$, respectively. Starting from the time $t > 5\tau _d^{\left( 2 \right)}$, there  practically establishes a stationary state of the distribution of the defects in a three-layer stressed nanoheterosystem. Thus, the deformation field of the interlayer ($- a \leqslant z \leqslant a $) clears away the workspace from the defects which finally makes the material of the workspace having a greater intensity of photoluminescence \cite{Che}. In exterlayers (figure~\ref{fig1}, regions~1, 3) of the heterosystem, the concentration of the defects asymmetrically monotonously increases from the boundary of the contacting materials and becomes larger than their average value $ N_{d0}^{\left( 1 \right)}$,  $ N_{d0}^{\left( 3 \right)}$.

If there are no defects ($ N_{d0}^{\left( i \right)}  =  0$) in the contacting materials, inhomogeneous deformation is created only due to a  mismatch between the lattice parameters of contacting materials [${\varepsilon _i}\left( z \right)  =  {\varepsilon _0} {{{{z^2}} \over {{a^2}}}}, i  =  2 $], and in the absence of the mismatch between the lattice parameters (${\varepsilon _0}  =  0 $), the deformation [$ {U^{\left( i \right)}}   =  ({{{\theta _d^{\left( i \right)}} / {{K^{\left( i \right)}}}}}) N_d^{\left( i \right)}$] is caused only by the point defects.

\section{Conclusions}
\begin{itemize}
  \item It has been established that the concentration profile of point defects at $ N_{d0}^{\left( i \right)} < N_{dc}^{\left(i\right)}$ is of nonmonotonous character with a minimum in the middle of the interlayer In$_x$Ga$_{1 - x}$As which is impoverished by the point defect of the type of compression centers when the interlayer of the heterostructure  GaAs/In$_x$Ga$_{1 - x}$As/GaAs undergoes an inhomogeneous compression, while in the case of inhomogeneous tension the opposite effect takes place. Thus, the deformation field and the number of the defects in the workspace of nanooptoelectronic devices can be controlled.
  \item It is shown that if the ratio of the thickness of the middle layer to the thicknesses of the external layers of a nanoheterosystem GaAs/In$_x$Ga$_{1 - x}$As/GaAs is ${{a}}/{L}=0.05$, the ratio of the initial average concentrations of the vacancies in the layers of a nanoheterosystem is  ${{{N_{d0}^{\left( 1 \right)}} / {N_{dc}^{\left( 1 \right)}}}}  = 0.5$, ${{{N_{d0}^{\left( 2 \right)}} / {N_{dc}^{\left( 2 \right)}}}}  = 0.8$, ${{{N_{d0}^{\left( 3 \right)}} / {N_{dc}^{\left( 3 \right)}}}}  = 0.6$ and the value of the deformation parameter is $\beta =10.2$, then the established concentration of vacancies in the middle layer ${Y}({\mathord{\buildrel{\lower3pt\hbox{$\scriptscriptstyle\frown$}}\over z}, 5\tau _d^{\left( 2 \right)}}) $ is less than the initial average concentration $N{}_{d0}^{\left( 2 \right)}$ by 16 \%. If the ratio ${{a}}/{L}=0.1$, then the established concentration of the  vacancies in the middle layer is larger than the initial average concentration $N_{d0}^{\left( 2 \right)}$. Such a reduction of the established concentration of the vacancies in the workspace of a nanoheterosystem is correlated with the experimental results of  the work \cite{Che}.

\end{itemize}


\section*{Appendix}

To find the solution of differential equations  (\ref{16}) with boundary conditions (\ref{17}), the integral Laplace transformation is used:
\begin{eqnarray}
\label{(A.1)}
{X_i}( {\mathord{\buildrel{\lower3pt\hbox{$\scriptscriptstyle\frown$}}
\over z} , p} )  =  \int\limits_0^\infty  {{Z_i}} ( {\mathord{\buildrel{\lower3pt\hbox{$\scriptscriptstyle\frown$}}
\over z} , \theta } ){\re^{ - p\theta }}\rd\theta.
\end{eqnarray}

Then, the differential equation (\ref{16}) and boundary conditions (\ref{17}) take up the form:
\begin{align}
&{H_1} X''_1\left( {\hat z,p} \right)  -  p {X_1}\left( {\hat z,p} \right)  + {{1 \over {p  -  1}}}  =  0,\nonumber\\
&{H_2} X''_2\left( {\hat z,p} \right)  -  p {X_2}\left( {\hat z,p} \right)  + {{{\beta   +  1} \over {p  -  1}}}  =  0,\label{(A.2)}\\
&{H_3} X''_3\left( {\hat z,p} \right)  -  p {X_3}\left( {\hat z,p} \right)  + {{1 \over {p  -  1}}}  =  0.\nonumber
\end{align}
\begin{equation}
\begin{split}
{X_1}\left( { - \mathord{\buildrel{\lower3pt\hbox{$\scriptscriptstyle\frown$}}
\over a} , p} \right)  &=  {X_2}\left( { - \mathord{\buildrel{\lower3pt\hbox{$\scriptscriptstyle\frown$}}
\over a} ,p} \right),\\
 -  {H_1}X_1^\prime \left( { - \mathord{\buildrel{\lower3pt\hbox{$\scriptscriptstyle\frown$}}
\over a} ,p} \right)  &=   -  {H_2}X_2^\prime \left( { - \mathord{\buildrel{\lower3pt\hbox{$\scriptscriptstyle\frown$}}
\over a} ,p} \right)  -  {{{\beta \mathord{\buildrel{\lower3pt\hbox{$\scriptscriptstyle\frown$}}
\over a} } \over {p - 1}}},\\
{X_2}\left( {\mathord{\buildrel{\lower3pt\hbox{$\scriptscriptstyle\frown$}}
\over a} , p} \right)  &=  {X_3}\left( {\mathord{\buildrel{\lower3pt\hbox{$\scriptscriptstyle\frown$}}
\over a} ,p} \right),\label{(A.3)}\\
-  {H_2}X_2^\prime \left( { \mathord{\buildrel{\lower3pt\hbox{$\scriptscriptstyle\frown$}}
\over a} ,p} \right)  +  {{{\beta \mathord{\buildrel{\lower3pt\hbox{$\scriptscriptstyle\frown$}}
\over a} } \over {p - 1}}} &=   -  {H_3}X_3^\prime \left( { \mathord{\buildrel{\lower3pt\hbox{$\scriptscriptstyle\frown$}}
\over a} ,p} \right),
\end{split}
\end{equation}
where ${H_i}  =  {1   -  {{{N_{d0}^{(i)}} / {N_{dc}^{(i)}}}}} $.

Analytical solutions of differential equations (\ref{(A.2)}) in each layer are as follows:
\begin{align}
&{X_1}( {\mathord{\buildrel{\lower3pt\hbox{$\scriptscriptstyle\frown$}}
\over z} ,p} )  =  {C_1}{\exp{\left(\sqrt {\frac{p}{{{H_1}}}\hat z} \right)}}
+ {{1 \over {p\left( {p  -  1} \right)}}}\,,&
- &\mathord{\buildrel{\lower3pt\hbox{$\scriptscriptstyle\frown$}}
\over a}  \leqslant  \mathord{\buildrel{\lower3pt\hbox{$\scriptscriptstyle\frown$}}
\over z}   \leqslant   -  \mathord{\buildrel{\lower3pt\hbox{$\scriptscriptstyle\frown$}}
\over l}\,, \nonumber\\
&{X_2}( {\mathord{\buildrel{\lower3pt\hbox{$\scriptscriptstyle\frown$}}
\over z} ,p} )  =  {C_2}{\exp\left({\sqrt {\frac{p}{{{H_2}}}} \hat z}\right)}  +   {C_3}{\exp\left({\sqrt {\frac{p}{{{H_2}}}} \hat z}\right)} +  {{{\beta  + 1} \over {p\left( {p  -  1} \right)}}}\,,&
- & \mathord{\buildrel{\lower3pt\hbox{$\scriptscriptstyle\frown$}}
\over a}  \leqslant  \mathord{\buildrel{\lower3pt\hbox{$\scriptscriptstyle\frown$}}
\over z}   \leqslant  \mathord{\buildrel{\lower3pt\hbox{$\scriptscriptstyle\frown$}}
\over a}\,,\label{(A.4)}
\\
&{X_3}( {\mathord{\buildrel{\lower3pt\hbox{$\scriptscriptstyle\frown$}}
\over z} ,p} )  =  {C_4}{\exp\left({{\sqrt {\frac{p}{{{H_2}}}} \hat z}}\right)} +  {{1 \over {p\left( {p  -  1} \right)}}}\,, & &\mathord{\buildrel{\lower3pt\hbox{$\scriptscriptstyle\frown$}}
\over a}  \leqslant  \mathord{\buildrel{\lower3pt\hbox{$\scriptscriptstyle\frown$}}
\over z}   \leqslant   -  \mathord{\buildrel{\lower3pt\hbox{$\scriptscriptstyle\frown$}}
\over l}. \nonumber
\end{align}

Integration constants ${C_1}$, ${C_2}$, ${C_3}$, ${C_4}$ are determined from the following system of algebraic equations:
\begin{equation}
\begin{split}
{C_1}{\exp\left( - \sqrt {{{p \over {{H_1}}}}} \mathord{\buildrel{\lower3pt\hbox{$\scriptscriptstyle\frown$}}
\over a} \right)}  -  {C_2}{\exp\left( - \sqrt {{{p \over {{H_2}}}}} \mathord{\buildrel{\lower3pt\hbox{$\scriptscriptstyle\frown$}}
\over a} \right)}  -  {C_3}{\exp\left(\sqrt {{{p \over {H_2}}}} \mathord{\buildrel{\lower3pt\hbox{$\scriptscriptstyle\frown$}}
\over a} \right)}  & =  {{\beta  \over {p\left( {p  -  1} \right)}}}\,,\\
- {C_1}\sqrt {{H_1}} {\exp\left( - \sqrt {{{p \over {{H_1}}}}} \mathord{\buildrel{\lower3pt\hbox{$\scriptscriptstyle\frown$}}
\over a} \right)}  +  {C_2}\sqrt {{H_2}} {\exp\left( - \sqrt {{{p \over {{H_2}}}}} \mathord{\buildrel{\lower3pt\hbox{$\scriptscriptstyle\frown$}}
\over a} \right)}  -  {C_3}\sqrt {{H_2}} {\exp\left(\sqrt {{{p \over {H2}}}} \mathord{\buildrel{\lower3pt\hbox{$\scriptscriptstyle\frown$}}
\over a} \right)}  & =  {{{\beta \mathord{\buildrel{\lower3pt\hbox{$\scriptscriptstyle\frown$}}
\over a} } \over {\sqrt p \left( {p  -  1} \right)}}}\,,
\label{(A.5)}\\
{C_2}{\exp\left(\sqrt {{{p \over {{H_2}}}}} \mathord{\buildrel{\lower3pt\hbox{$\scriptscriptstyle\frown$}}
\over a} \right)}  +  {C_3}{\exp\left( - \sqrt {{{p \over {{H_2}}}}} \mathord{\buildrel{\lower3pt\hbox{$\scriptscriptstyle\frown$}}
\over a} \right)}  -  {C_4}{\exp\left( - \sqrt {{{p \over {{H_3}}}}} \mathord{\buildrel{\lower3pt\hbox{$\scriptscriptstyle\frown$}}
\over a} \right)} & =   -   {{\beta  \over {p\left( {p  -  1} \right)}}}\,,\\
- {C_2}\sqrt {{H_2}} {\exp\left(\sqrt {{{p \over {H2}}}} \mathord{\buildrel{\lower3pt\hbox{$\scriptscriptstyle\frown$}}
\over a} \right)}  +  {C_3}\sqrt {{H_2}} {\exp\left( - \sqrt {{{p \over {{H_2}}}}} \mathord{\buildrel{\lower3pt\hbox{$\scriptscriptstyle\frown$}}
\over a} \right)}  +  {C_4}\sqrt {{H_3}} {\exp\left( - \sqrt {{{p \over {H3}}}} \mathord{\buildrel{\lower3pt\hbox{$\scriptscriptstyle\frown$}}
\over a} \right)}  & =  {{{\beta \mathord{\buildrel{\lower3pt\hbox{$\scriptscriptstyle\frown$}}
\over a} } \over {\sqrt p \left( {p  -  1} \right)}}}\, .
\end{split}
\end{equation}

By carrying out the inverse Laplace transformation we obtain the spatial-temporal redistribution of the point defects in the first, second and third stressed layers,  respectively:
\begin{align}
{Y_1}( {\mathord{\buildrel{\lower3pt\hbox{$\scriptscriptstyle\frown$}}
\over z} ,\theta } )  &=  {\re^{ - \theta }}{Z_1}( {\mathord{\buildrel{\lower3pt\hbox{$\scriptscriptstyle\frown$}}
\over z} ,\theta } )  -  {\re^{ - \theta }}  +  1, \nonumber \\
{Y_2}( {\mathord{\buildrel{\lower3pt\hbox{$\scriptscriptstyle\frown$}}
\over z} ,\theta } )  &=  {Z_2}( {\mathord{\buildrel{\lower3pt\hbox{$\scriptscriptstyle\frown$}}
\over z} ,\theta } ) \left( {1 - {\re^{ - \theta }}} \right)\left( {1 + \beta } \right),
\label{(A.6)}\\
{Y_3}( {\mathord{\buildrel{\lower3pt\hbox{$\scriptscriptstyle\frown$}}
\over z} ,\theta } )  &=  {\re^{ - \theta }}{Z_3}( {\mathord{\buildrel{\lower3pt\hbox{$\scriptscriptstyle\frown$}}
\over z} ,\theta } )  -  {\re^{ - \theta }}  +  1,\nonumber
\end{align}
where functions ${Z_1}( {\mathord{\buildrel{\lower3pt\hbox{$\scriptscriptstyle\frown$}}
\over z} ,\theta })$, ${Z_2}( {\mathord{\buildrel{\lower3pt\hbox{$\scriptscriptstyle\frown$}}
\over z} ,\theta } )$ and ${Z_3}( {\mathord{\buildrel{\lower3pt\hbox{$\scriptscriptstyle\frown$}}
\over z} ,\theta } )$ are the solutions of differential equations  (\ref{16}) with boundary conditions  (\ref{17}), respectively:
\begin{eqnarray}{Z_1}( {\mathord{\buildrel{\lower3pt\hbox{$\scriptscriptstyle\frown$}}
\over z} ,\theta } ) & = & {{1 \over b}}\left[ {{\re^\theta }  -  1  + {\re^\theta }{{{\mathord{\buildrel{\lower3pt\hbox{$\scriptscriptstyle\frown$}}
\over a} } \over {\sqrt {{H_2}} }}}\textrm{Erf}\left( {\sqrt \theta  } \right)} \right]{l_3}
+ {\sum\limits_{i = 1}^2 {\sum\limits_{m = 0}^\infty ({-1})^m }}{{{{a^m}} \over {{b^{m + 1}}}}}
\nonumber\\
&\times&\sum\limits_{k = 1}^\infty  {{l_{2i - 1}}} {r_{2i - 1}}{\theta ^k}{\Phi _1}(\mathord{\buildrel{\lower3pt\hbox{$\scriptscriptstyle\frown$}}
\over z} ,\theta ,k)+{\sum\limits_{m = 0}^\infty  {\left( { - 1} \right)} ^m}{{{{a^m}} \over {{b^{m + 1}}}}}\sum\limits_{k = 1}^\infty  {{l_2}} {r_2}{\theta ^k}{\Phi _2}(\mathord{\buildrel{\lower3pt\hbox{$\scriptscriptstyle\frown$}}
\over z} ,\theta ,k),
\end{eqnarray}
where
\begin{eqnarray}
{\Phi _1}(\mathord{\buildrel{\lower3pt\hbox{$\scriptscriptstyle\frown$}}
\over z} ,\theta ,k) &=&  {{{{}_1{F_1}\left( {{{1 \over 2}} - k,{{3 \over 2}} ,  -  {{{{r_{2i - 1}^2}} \over {4\theta }}}} \right)} \over {\sqrt \theta   \Gamma \left( {k  + {{1 \over 2}}} \right)}}}   + {{{{}_1{F_1}\left( { - k,{{1 \over 2}} ,  -  {{{{r_{2i - 1}^2}} \over {4\theta }}}} \right)} \over {{r_{2i - 1}} \Gamma \left( {k  + 1} \right)}}} \nonumber\\
&-&  {\left( { -  1} \right)^i}{{{{}_1{F_1}\left( {{{1 \over 2}} - k,{{1 \over 2}} ,  -  {{{{r_{2i - 1}^2}} \over {4\theta }}}} \right)} \over {\sqrt \theta   {r_{2i - 1}}\Gamma \left( {k  + {{1 \over 2}}} \right)}}}  + {{{{}_1{F_1}\left( {1 - k,{{3 \over 2}} ,  -  {{{{r_{2i - 1}^2}} \over {4\theta }}}} \right)} \over {\theta  \Gamma \left( {k  + {{1 \over 2}}} \right)}}}\,,
\end{eqnarray}
\begin{eqnarray}{\Phi _2}(\mathord{\buildrel{\lower3pt\hbox{$\scriptscriptstyle\frown$}}
\over z} ,\theta ,k) = -   {{{{}_1{F_1}\left( {{{1 \over 2}} - k,{{3 \over 2}} ,  -  {{{{r_2^2}} \over {4\theta }}}} \right)} \over {\sqrt \theta   \Gamma \left( {k  + {{1 \over 2}}} \right)}}}   + {{{{}_1{F_1}\left( { - k,{{1 \over 2}} ,  -  {{{{r_2^2}} \over {4\theta }}}} \right)} \over {{r_2} \Gamma \left( {k  + 1} \right)}}} \,,
\end{eqnarray}
\[a  =  \left( {\sqrt {{H_1}}   -  \sqrt {{H_2}} } \right)\left( {\sqrt {{H_2}}   -  \sqrt {{H_3}} } \right), \qquad
b  =  \left( {\sqrt {{H_1}}   +  \sqrt {{H_2}} } \right)\left( {\sqrt {{H_2}}   +  \sqrt {{H_3}} } \right)\,,\]
\[{l_1}  =  \beta \sqrt {{H_2}} \left( {\sqrt {{H_3}}   -  \sqrt {{H_2}} } \right),
\qquad {l_2}  = -  {{{{l_1}\mathord{\buildrel{\lower3pt\hbox{$\scriptscriptstyle\frown$}}
\over a} } \over {\sqrt {{H_2}} }}}\,,\]
\[{l_3}  =  \beta \sqrt {{H_2}} \left( {\sqrt {{H_3}}   +  \sqrt {{H_2}} } \right),
\qquad {l_4}  = {{{{l_3}\mathord{\buildrel{\lower3pt\hbox{$\scriptscriptstyle\frown$}}
\over a} } \over {\sqrt {{H_2}} }}}\,,\]
\[{r_1}  =   - {{{\mathord{\buildrel{\lower3pt\hbox{$\scriptscriptstyle\frown$}}
\over z}   +  \mathord{\buildrel{\lower3pt\hbox{$\scriptscriptstyle\frown$}}
\over a} } \over {\sqrt {{H_1}} }}}  + {{{4 \mathord{\buildrel{\lower3pt\hbox{$\scriptscriptstyle\frown$}}
\over a} \left( {m + 1} \right)} \over {\sqrt {{H_2}} }}}\,,\qquad
{r_2}  =   - {{{\mathord{\buildrel{\lower3pt\hbox{$\scriptscriptstyle\frown$}}
\over z}   +  \mathord{\buildrel{\lower3pt\hbox{$\scriptscriptstyle\frown$}}
\over a} } \over {\sqrt {{H_1}} }}}  + {{{2 \mathord{\buildrel{\lower3pt\hbox{$\scriptscriptstyle\frown$}}
\over a} \left( {2m + 1} \right)} \over {\sqrt {{H_2}} }}}\,,
\qquad
{r_3}  =   - {{{\mathord{\buildrel{\lower3pt\hbox{$\scriptscriptstyle\frown$}}
\over z}   +  \mathord{\buildrel{\lower3pt\hbox{$\scriptscriptstyle\frown$}}
\over a} } \over {\sqrt {{H_1}} }}}  + {{{4 \mathord{\buildrel{\lower3pt\hbox{$\scriptscriptstyle\frown$}}
\over a} m} \over {\sqrt {{H_2}} }}}\,.\]
\begin{eqnarray}
{Z_2}( {\mathord{\buildrel{\lower3pt\hbox{$\scriptscriptstyle\frown$}}
\over z} ,\theta } )  &=&  {{1 \over b}}\left[ {({\re^\theta }  -  1 )({p_3}  + {p_4}) + ({p_5}  +   {p_6}) {\re^\theta }\textrm{Erf}( {\sqrt \theta  } )} \right] \nonumber  \\
&+&{\sum\limits_{i = 1}^4 {\sum\limits_{m = 0}^\infty  {\left( { - 1} \right)^m} } }{{{{a^m}} \over {{b^{m + 1}}}}}\sum\limits_{k = 1}^\infty  {{p_i}} {e_i}{\theta ^k}{\Phi _3}( {\mathord{\buildrel{\lower3pt\hbox{$\scriptscriptstyle\frown$}}
\over z} ,\theta ,k} ),
\end{eqnarray}
where
\begin{eqnarray}
{\Phi _3}(\mathord{\buildrel{\lower3pt\hbox{$\scriptscriptstyle\frown$}}
\over z} ,\theta ,k) &=&  -   {{{{}_1{F_1}\left( {{{1 \over 2}} - k,{{3 \over 2}} ,  -  {{{{e_i^2}} \over {4\theta }}}} \right)} \over {\sqrt \theta   \Gamma \left( {k  + {{1 \over 2}}} \right)}}}   + {{{{}_1{F_1}\left( { - k,{{1 \over 2}} ,  -  {{{{e_i^2}} \over {4\theta }}}} \right)} \over {{e_i} \Gamma \left( {k  + 1} \right)}}} \nonumber\\
&+&   {{{\mathord{\buildrel{\lower3pt\hbox{$\scriptscriptstyle\frown$}}
\over a} } \over {\sqrt {{H_2}} }}}{{{{}_1{F_1}\left( {{{1 \over 2}} - k,{{1 \over 2}} ,  -  {{{e_i^2} \over {4\theta }}}} \right)} \over {\sqrt \theta   {e_i}\Gamma \left( {k  + {{1 \over 2}}} \right)}}}  + {{{{}_1{F_1}\left( {1 - k,{{3 \over 2}} ,  -  {{{e_i^2} \over {4\theta }}}} \right)} \over {\theta  \Gamma \left( {k  + {{1 \over 2}}} \right)}}}\,,
\end{eqnarray}
\[{p_1}  =  \beta \sqrt {{H_2}} \left( {\sqrt {{H_3}}   -  \sqrt {{H_2}} } \right),
\qquad
{p_2}  = \beta \mathord{\buildrel{\lower3pt\hbox{$\scriptscriptstyle\frown$}}
\over a} \left( {\sqrt {{H_2}}   -  \sqrt {{H_3}} } \right),
\qquad
{p_3}  =   - {{{{p_2}\mathord{\buildrel{\lower3pt\hbox{$\scriptscriptstyle\frown$}}
\over a} } \over {\sqrt {{H_2}} }}}\,,
\]
\[{p_4}  =   - \beta \sqrt {{H_3}} \left( {\sqrt {{H_1}}   +  \sqrt {{H_2}} } \right),
\qquad
{p_5}  =   -  {{{{p_4}\mathord{\buildrel{\lower3pt\hbox{$\scriptscriptstyle\frown$}}
\over a} } \over {\sqrt {{H_3}} }}}\,,
\qquad
{p_6}  =  \beta \sqrt {{H_1}} \left( {\sqrt {{H_2}}   - \sqrt {{H_3}} } \right),\]
\[{e_1}  =   - {{{\mathord{\buildrel{\lower3pt\hbox{$\scriptscriptstyle\frown$}}
\over z}   +  \mathord{\buildrel{\lower3pt\hbox{$\scriptscriptstyle\frown$}}
\over a} } \over {\sqrt {{H_2}} }}}  + {{{4 \mathord{\buildrel{\lower3pt\hbox{$\scriptscriptstyle\frown$}}
\over a} \left( {m + 1} \right)} \over {\sqrt {{H_2}} }}}\,, \qquad
{e_2}  =   - {{{\mathord{\buildrel{\lower3pt\hbox{$\scriptscriptstyle\frown$}}
\over z}   +  \mathord{\buildrel{\lower3pt\hbox{$\scriptscriptstyle\frown$}}
\over a} } \over {\sqrt {{H_2}} }}}  + {{{2 \mathord{\buildrel{\lower3pt\hbox{$\scriptscriptstyle\frown$}}
\over a} \left( {2m + 1} \right)} \over {\sqrt {{H_2}} }}},
\]
\[
{e_3}  =  {{{\mathord{\buildrel{\lower3pt\hbox{$\scriptscriptstyle\frown$}}
\over z}   -  \mathord{\buildrel{\lower3pt\hbox{$\scriptscriptstyle\frown$}}
\over a} } \over {\sqrt {{H_2}} }}}  + {{{4 \mathord{\buildrel{\lower3pt\hbox{$\scriptscriptstyle\frown$}}
\over a} \left( {m + 1} \right)} \over {\sqrt {{H_2}} }}}\,,
\qquad
{e_4}  = {{{\mathord{\buildrel{\lower3pt\hbox{$\scriptscriptstyle\frown$}}
\over z}   -  \mathord{\buildrel{\lower3pt\hbox{$\scriptscriptstyle\frown$}}
\over a} } \over {\sqrt {{H_2}} }}}  + {{{2 \mathord{\buildrel{\lower3pt\hbox{$\scriptscriptstyle\frown$}}
\over a} \left( {2m + 1} \right)} \over {\sqrt {{H_2}} }}}\,.
\]
\begin{eqnarray}
{Z_3}( {\mathord{\buildrel{\lower3pt\hbox{$\scriptscriptstyle\frown$}}
\over z} ,\theta } )  &=&  {{1 \over b}}\left[ {{\re^\theta }  -  1  + {\re^\theta }{{{\mathord{\buildrel{\lower3pt\hbox{$\scriptscriptstyle\frown$}}
\over a} } \over {\sqrt {{H_2}} }}}\textrm{Erf}( {\sqrt \theta  } )} \right]{l_3}  + {\sum\limits_{i = 1}^2 {\sum\limits_{m = 0}^\infty  {\left( { - 1} \right)^m} } }{{{{a^m}} \over {{b^{m + 1}}}}}
\nonumber\\
&\times&
\sum\limits_{k = 1}^\infty  {{l_{2i - 1}}} {r_{2i - 1}}{\theta ^k}{\Phi _4}( {\mathord{\buildrel{\lower3pt\hbox{$\scriptscriptstyle\frown$}}
\over z} ,\theta ,k} )  +  {\sum\limits_{m = 0}^\infty  {\left( { - 1} \right)} ^m}{{{{a^m}} \over {{b^{m + 1}}}}}\sum\limits_{k = 1}^\infty  {{l_2}} {r_2}{\theta ^k}{\Phi _5}( {\mathord{\buildrel{\lower3pt\hbox{$\scriptscriptstyle\frown$}}
\over z} ,\theta ,k} ),
\end{eqnarray}
where
\begin{eqnarray}
{\Phi _4}(\mathord{\buildrel{\lower3pt\hbox{$\scriptscriptstyle\frown$}}
\over z} ,\theta ,k) &=&  -   {{{{}_1{F_1}\left( {{{1 \over 2}} - k,{{3 \over 2}} ,  -  {{{{r_{2i - 1}^2}} \over {4\theta }}}} \right)} \over {\sqrt \theta   \Gamma \left( {k  + {{1 \over 2}}} \right)}}}   + {{{{}_1{F_1}\left( { - k,{{1 \over 2}} ,  -  {{{{r_{2i - 1}^2}} \over {4\theta }}}} \right)} \over {{r_{2i - 1}} \Gamma \left( {k  + 1} \right)}}}  \nonumber\\
&+&   {{{\mathord{\buildrel{\lower3pt\hbox{$\scriptscriptstyle\frown$}}
\over a} } \over {\sqrt {{H_2}} }}}{{{{}_1{F_1}\left( {{{1 \over 2}} - k,{{1 \over 2}} ,  -  {{{{r_{2i - 1}^2}} \over {4\theta }}}} \right)} \over {\sqrt \theta   {r_{2i - 1}}\Gamma \left( {k  + {{1 \over 2}}} \right)}}}  + {{{{}_1{F_1}\left( {1 - k,{{3 \over 2}} ,  -  {{{{r_{2i - 1}^2}} \over {4\theta }}}} \right)} \over {\theta  \Gamma \left( {k  + {{1 \over 2}}} \right)}}}\,,
\end{eqnarray}
\begin{eqnarray}
{\Phi _5}(\mathord{\buildrel{\lower3pt\hbox{$\scriptscriptstyle\frown$}}
\over z} ,\theta ,k) = -   {{{{}_1{F_1}\left( {{{1 \over 2}} - k,{{3 \over 2}} ,  -  {{{{r_2^2}} \over {4\theta }}}} \right)} \over {\sqrt \theta   \Gamma \left( {k  + {{1 \over 2}}} \right)}}}   + {{{{}_1{F_1}\left( { - k,{{1 \over 2}} ,  -  {{{{r_2^2}} \over {4\theta }}}} \right)} \over {{r_2} \Gamma \left( {k  + 1} \right)}}} \,,
\end{eqnarray}
\[{l_1}  =  \beta \sqrt {{H_2}} \left( {\sqrt {{H_1}}   -  \sqrt {{H_2}} } \right),\qquad
{l_2}  = -  {{{{l_1}\mathord{\buildrel{\lower3pt\hbox{$\scriptscriptstyle\frown$}}
\over a} } \over {\sqrt {{H_2}} }}}\,,\]
\[{l_3}  =  \beta \sqrt {{H_2}} \left( {\sqrt {{H_2}}   +  \sqrt {{H_1}} } \right),
\qquad
{l_4}  = {{{{l_3}\mathord{\buildrel{\lower3pt\hbox{$\scriptscriptstyle\frown$}}
\over a} } \over {\sqrt {{H_2}} }}}\,,\]
\[{r_1}  =   {{{\mathord{\buildrel{\lower3pt\hbox{$\scriptscriptstyle\frown$}}
\over z}   -  \mathord{\buildrel{\lower3pt\hbox{$\scriptscriptstyle\frown$}}
\over a} } \over {\sqrt {{H_1}} }}}  + {{{4 \mathord{\buildrel{\lower3pt\hbox{$\scriptscriptstyle\frown$}}
\over a} \left( {m + 1} \right)} \over {\sqrt {{H_2}} }}}\,,
\qquad
{r_2}  = {{{\mathord{\buildrel{\lower3pt\hbox{$\scriptscriptstyle\frown$}}
\over z}   -  \mathord{\buildrel{\lower3pt\hbox{$\scriptscriptstyle\frown$}}
\over a} } \over {\sqrt {{H_1}} }}}  + {{{2 \mathord{\buildrel{\lower3pt\hbox{$\scriptscriptstyle\frown$}}
\over a} \left( {2m + 1} \right)} \over {\sqrt {{H_2}} }}}\,,
\qquad
{r_3}  = {{{\mathord{\buildrel{\lower3pt\hbox{$\scriptscriptstyle\frown$}}
\over z}   -  \mathord{\buildrel{\lower3pt\hbox{$\scriptscriptstyle\frown$}}
\over a} } \over {\sqrt {{H_1}} }}}  + {{{4 \mathord{\buildrel{\lower3pt\hbox{$\scriptscriptstyle\frown$}}
\over a} m} \over {\sqrt {{H_2}} }}}\,.\]

\ukrainianpart

\title{Просторово-часовий перерозподіл точкових дефектів у тришарових напружених  наногетеросистемах  у межах самоузгодженої деформаційно-дифузійної моделі}
\author{Р.М.~Пелещак, Н.Я.~Кулик, М.В.~Дорошенко}
\address{Дрогобицький державний педагогічний університет ім.~Івана Франка,\\
вул. І.~Франка, 24, 82100 Дрогобич, Україна}

\makeukrtitle

\begin{abstract}
\tolerance=3000%
Побудовано модель просторово-часового розподілу точкових дефектів у тришаровій напруженій наногетеросистемі GaAs/In$_x$Ga$_{1 - x}$As/GaAs з врахуванням самоузгодженої деформаційно-дифузійної взаємодії. У межах цієї моделі розраховано профіль просторово-часового розподілу вакансій (міжвузлових атомів) у напруженій наногетеросистемі GaAs/In$_x$Ga$_{1 - x}$As/GaAs та показано, що у випадку встановлення стаціонарного стану ($ t  >5\tau _d^{\left( 2 \right)}$), концентрація вакансій в неоднорідно стиснутому внутрішньому шарі  є меншою відносно вихідного середнього значення $N_{d0}^{(2)}$ на 16\%.
\keywords  просторово-часовий перерозподіл, вакансії, міжвузлові атоми
\end{abstract}

\end{document}